\documentclass[12pt]{revtex4}
\usepackage{amssymb}
\usepackage{latexsym}
\usepackage{epsfig}
\begin{document}                             

\title{ Emergence and expansion of cosmic space as due to M0-branes
 }
\author{Alireza Sepehri $^{1,2}$\footnote{alireza.sepehri@uk.ac.ir}, Mohammad Reza Setare $^{3}$\footnote{rezakord@ipm.ir}, Salvatore Capozziello $^{4,5,6}$
\footnote{capozziello@na.infn.it} }
\address{$^1$Faculty of Physics,
Shahid Bahonar University, P.O. Box 76175, Kerman, Iran.\\$^{2}$
Research Institute for Astronomy and Astrophysics of Maragha
(RIAAM), P.O. Box 55134-441, Maragha, Iran.\\ $^3$ Department of
Science, Campus of Bijar, University of Kurdistan, Bijar, Iran.
\\{$^{4 }$ Dipartimento di Fisica, Universit´a di Napoli
Federico II, I-80126 - Napoli,
Italy. \\
$^{ 5 }$ INFN Sez. di Napoli, Compl. Univ. di Monte S. Angelo,
Edificio G, I-80126 - Napoli, Italy,\\ $^{  6}$ Gran Sasso Science
Institute (INFN), Viale F. Crispi, 7, I-67100, LAquila, Italy} }

\begin{abstract}
Recently, Padmanabhan [arXiv:1206.4916 [hep-th]]discussed that the difference
between the number of degrees of freedom on the boundary surface and the number of degrees
of freedom in a bulk region causes to the accelerated expansion of the universe. The main question arises as to what is the origin of this inequality between the surface degrees of freedom and the bulk degrees of freedom? We answer this question in M-theory. In our model, first M0-branes are compactified on one circle and N D0-branes are created. Then,  N D0-branes join to each other, grow and form one D5-branes. Next, D5- brane is compactified on two circle and our universe-D3-brane, two D1-brane and some extra energies are produced. After that, one of D1-branes which is more close to universe-brane, gives it's energy into it, leads to an increase in difference between number of degrees of freedom and occurring inflation era. With the disappearance of this D1-brane, the number of degrees of freedom of boundary surface and bulk region become equal and inflation ends. At this stage, extra energies that are produced due to the compactification cause to an expansion of universe and deceleration epoch.  Finally, another D1-brane, dissolves in our universe-brane , leads to an inequality between degrees of freedom and happening a new phase of acceleration.

\end{abstract}

 \maketitle
\section{Introduction}
About three years ago,  Padmanabhan suggested that the accelerated expansion of the universe is due to the difference
between the surface degrees of freedom on the holographic horizon and the bulk degrees of freedom through a simple equation
$\triangle V = \triangle t(N_{sur} - N_{bulk})$ where V is the Hubble volume in Planck units and t is the cosmic time
in Planck units \cite{q1}. Up to date, many discussions have been done on the Padmanabhan proposal \cite{q2,q3,q4,q5,q6,q7,q8}. For example, in one paper, with the help of this idea, the Friedmann equations of an (n + 1)-dimensional
Friedmann-Robertson-Walker universe corresponding to general relativity, Gauss-Bonnet gravity, and Lovelock gravity have been obtained  \cite{q2}. In another research, the idea of treating the cosmic space as an emergent process has been applied to brane cosmology, scalar-tensor cosmology, and f(R) gravity and the corresponding
cosmological equations in these theories have been derived \cite{q3}. In another investigation, using Padmanabhan suggestion, author obtained the Friedmann equations of universe not only in four dimensional space-time and  Einstein gravity, but also in higher dimensional space-time and  other gravities like Gauss-Bonnet and
 Lovelock gravity with any spacial curvature \cite{q4}. Some other authors, have extended the evolution
equation in Padmanabhan idea to give the Friedmann equation  in
the nonflat universe corresponding to $k = \pm 1$ by taking into
account the invariant volume surrounded by the apparent horizon
\cite{q5}. In another scenario, authors showed that applying
Padmanabhan’s conjecture to non-Einstein gravity cases
encounters serious difficulties and  has to be heavily modified to
get the Friedmann equation \cite{q6}. In another paper, author
applied  derived equations of universe in the Padmanabhan model
with the corrected entropy-area law that follows from Generalized
Uncertainty Principle (GUP) and obtain a modified Friedmann
equations due to the GUP \cite{q7}. and in more recent research,
the Padmanabhan idea has been constructed in BIonic system and
shown that all degrees of freedom inside and outside the universe
are controlled by the evolutions of BIon in extra dimension and
tend to degrees of freedom of black F-string in string theory
\cite{q8}. The BIon is a configuration of two branes which are
connected by a wormhole \cite{E1,E2,E3,E4}.

Now, the main question arises as to what is the origin of difference between the number of degrees of freedom on the boundary surface and the one
 in a bulk?  To answer this question, we use the method in \cite{E4}. In that
 paper, k fundamental strings decay to N pairs of M0-branes. Then,
 these branes glue to each other and form an M3, an
 anti-M3 and a wormhole between them. Our universe is located on
 one of these branes and interact with other brane via the
 wormhole.  Extending this idea , we propose a new model which allows to construct our universe from M0-branes in M-theory.
  In this proposal, first, we will compactify M0-branes  on one circle and obtain the relevant action for N D0-branes.
   Then, we will show that these D0-branes may join to each other and make a transition to a D5-brane.
    Next, we will compactify this D5- brane  on two circle and derive the relevant actions for one D3-brane,
     two D1-branes and some extra energies. Our universe is located on this D3-brane and two D1-branes are
      the main causes of inequality between number of degrees freedom on the surface horizon and in a bulk and occurring inflation
       and late-time acceleration. Also, extra energies are responsible for happening of deceleration epoch.

The outline of the
paper is as  follows.  In section \ref{o1}, we will construct D0-brane from M0-brane and consider the relation between their algebra. We also show that D0-branes can join each other and form a D5-brane . Then, we will compactify D5-brane on two circles and obtain one D3-brane and two D1-branes. In section
\ref{o2}, we will show that the D1-brane which is more close to D3-brane dissolves in it and leads to  inflation.  Also, in this section
, we will argue that another D1-brane is the main cause of second phase of acceleration and inequality between number of degrees of freedom on the holographic surface and one in a bulk .  The last section is
devoted to summary and conclusion.

\section{ The birth of universe in M-theory}\label{o1}
In this section, we will show that the origin of universe is
M0-branes. Recently, some authors have proposed an action based on
Lie 3-algebras to describe M2-branes \cite{q9,q10,q11,q12,q13}.
Some other authors have considered the case of infinite
dimensional Lie 3-algebras based on the Nambu-Poisson structure of
three dimensional manifolds. They argued that the model contains
self-dual 2-form gauge fields in 6 dimensions, and the result may
be interpreted as the M5-brane world-volume action \cite{q14}.
Extending these methods, we propose a new model which allows to
construct our universe-brane from M0-branes. To this end, first,
we obtain the relevant action for N M0-branes by replacing
Nambu-Poisson structure of two dimensional manifolds in D-branes
by the structure of three dimensional one. At second stage, we
will compactify them on one circle and derive the action for N
D0-brane. We show that N D0-branes join to each other, grow and
form a D5-brane. Then, this brane is compactified on two circles
and our universe-brane and two D1-branes are created.

First, we introduce the Born-Infeld action for M0-brane by
replacing two dimensional Nambu-Poisson bracket \cite{
q15,q16,q17,q18,q19,q20} in the action of Dp-branes by three
dimensional Nambu-Poisson bracket \cite{q9,q10,q11,q12,q13,q14}
and applying Lie 3-algebra \cite{E4}.

\begin{eqnarray}
S_{M0} =
T_{M0}\int dt Tr(
\Sigma_{M,N,L=0}^{10}
\langle[X^{M},X^{N},X^{L}],[X^{M},X^{N},X^{L}]\rangle)
\label{a1}
\end{eqnarray}

where $X^{M}=X^{M}_{\alpha}T^{\alpha}$ and

\begin{eqnarray}
 &&[T^{\alpha}, T^{\beta}, T^{\gamma}]= f^{\alpha \beta \gamma}_{\eta}T^{\eta} \nonumber \\&&\langle T^{\alpha}, T^{\beta} \rangle = h^{\alpha\beta} \nonumber \\&& [X^{M},X^{N},X^{L}]=[X^{M}_{\alpha}T^{\alpha},X^{N}_{\beta}T^{\beta},X^{L}_{\gamma}T^{\gamma}]\nonumber \\&&\langle X^{M},X^{M}\rangle = X^{M}_{\alpha}X^{M}_{\beta}\langle T^{\alpha}, T^{\beta} \rangle
\label{a2}
\end{eqnarray}

where  $X^{M}$(i=1,3,...10)'s refer to transverse scalars to
M0-brane. One can show that by compactifying M-theory on a circle
of radius R,
 the above action transits to ten dimensional action for D0-brane \cite{q18,q20}.
  To this end, we apply the method in \cite{q21} and define $<X^{10}>=\frac{R}{l_{p}^{3/2}}$ where $l_{p}$ is the Planck length. We obtain \cite{E4}:

\begin{eqnarray}
&& S_{M0} = -
T_{M0}\int dt Tr(
\Sigma_{M,N,L=0}^{10}
\langle[X^{M},X^{N},X^{L}],[X^{M},X^{N},X^{L}]\rangle) = \nonumber \\
&& - T_{M0}\int dt Tr(\Sigma_{M,N,L,E,F,G=0}^{10}\varepsilon_{MNLD}\varepsilon_{EFG}^{D}X^{M}X^{N}X^{L}X^{E}X^{F}X^{G} = \nonumber \\
&& - 6T_{M0}\int dt Tr(\Sigma_{M,N,E,F=0}^{9}\varepsilon_{MN10D}\varepsilon_{EF10}^{D}X^{M}X^{N}X^{10}X^{E}X^{F}X^{10} - \nonumber \\
&& 6T_{M0} \int dt \Sigma_{M,N,L,E,F,G=0,\neq 10}^{9}\varepsilon_{MNLD}\varepsilon_{EFG}^{D}X^{M}X^{N}X^{L}X^{E}X^{F}X^{G} = \nonumber \\
&& - 6T_{M0}(\frac{R^{2}}{l_{p}^{3}})\int dt Tr(\Sigma_{M,N,E,F=0}^{9}\varepsilon_{MN10D}\varepsilon_{EF10}^{D}X^{M}X^{N}X^{E}X^{F} - \nonumber \\
&& 6T_{M0}\int dt \Sigma_{M,N,L,E,F,G=0,\neq10}^{9}\varepsilon_{MNLD}\varepsilon_{EFG}^{D}X^{M}X^{N}X^{L}X^{E}X^{F}X^{G} = \nonumber \\
&& - 6T_{M0}(\frac{R^{2}}{l_{p}^{3}})\int dt Tr(\Sigma_{M,N=0}^{9}[X^{M},X^{N}]^{2}) - \nonumber \\
&& 6T_{M0} \int dt \Sigma_{M,N,L,E,F,G=0,\neq 10}^{9}\varepsilon_{MNLD}\varepsilon_{EFG}^{D}X^{M}X^{N}X^{L}X^{E}X^{F}X^{G}= \nonumber \\
&& S_{D0} -  6T_{M0} \int dt \Sigma_{M,N,L,E,F,G=0,\neq10}^{9}\varepsilon_{MNLD}\varepsilon_{EFG}^{D}X^{M}X^{N}X^{L}X^{E}X^{F}X^{G}\nonumber \\
&& S_{D0} + V_{Extra,1}
\label{a2}
\end{eqnarray}

where   $ V_{Extra,1}= -6T_{M0}\int dt
\Sigma_{M,N,L,E,F,G=0}^{9}\varepsilon_{MNLD}\varepsilon_{EFG}^{D}X^{M}X^{N}X^{L}X^{E}X^{F}X^{G}$.
 Also, $T_{M0}$ and $T_{D0}$ denote tensions of M0 and D0 respectively and $T_{D0}=6T_{M0}(\frac{R^{2}}{l_{p}^{3}})=\frac{1}{g_{s}l_{s}}$ where $g_{s}$ and $l_{s}$ are the string coupling and string length respectively.
    Clearly, the action (\ref{a2}) for compactified M0-branes  is equal to the  sum of relevant action for D0-brane \cite{E4,q15,q16,q17,q18,q19,q20}

\begin{eqnarray}
&& S_{D0} =
-T_{D0}
\int dt Tr(
\Sigma_{m=0}^{9}
[X^{m},X^{n}]^{2})
\label{a3}
\end{eqnarray}

and some extra energies that are produced due to compactification. Now, we can construct other Dp-branes from D0-branes by substituting
 following rules \cite{E4,q15,q16,q17,q18,q19,q20}:

\begin{eqnarray}
&& \Sigma_{a=0}^{p}\Sigma_{m=0}^{9}\rightarrow \frac{1}{(2\pi l_{s})^{p}}\int d^{p+1}\sigma \Sigma_{m=p+1}^{9}\Sigma_{a=0}^{p} \qquad \lambda = 2\pi l_{s}^{2} \nonumber \\
&&[X^{a},X^{i}]=i
\lambda \partial_{a}X^{i}\qquad  [X^{a},X^{b}]= i \lambda^{2} F^{ab}\nonumber \\
&& i,j=p+1,..,9\qquad a,b=0,1,...p\qquad m,n=0,1,..,9
\label{a4}
\end{eqnarray}

in action (\ref{a3}) and doing some mathematical calculations
\cite{E4}:

\begin{eqnarray}
&& S_{Dp} = \Sigma_{a=0}^{p}S_{D0}=-\Sigma_{a=0}^{p}T_{D0}
\int dt Tr(
\Sigma_{m=0}^{9}
[X^{m},X^{n}]^{2}) =  \nonumber \\ &&
-T_{Dp} \int d^{p+1}\sigma Tr
(\Sigma_{a,b=0}^{p}
\Sigma_{i,j=p+1}^{9}
\{\partial_{a}X^{i}\partial_{b}X^{i}-\frac{1}{2 \lambda^{2}}[X^{i},X^{j}]^{2}+\frac{\lambda^{2}}{4}
(F_{ab})^{2}
\})
\label{a5}
\end{eqnarray}

which is in agreement with
results of \cite{q15,q16,q17,q18,q19,q20} for D1, D3 and D5-branes.
 Here $T_{Dp}=\frac{T_{D0}}{(2\pi l_{s})^{p}}$ is the brane tension,
  $l_{s}$ is the string length, $g_{s}$ is the string coupling and $F_{ab}$ is the field strength.
   To compactify this Dp-brane on one circle, we need to replace some gauge fields by some scalar fields use of following laws
     \cite{E4,q14,q15,q16,q17,q18,q19,q20,q21}:

\begin{eqnarray}
&& T_{Dp} \int d^{p+1}\sigma Tr
\Sigma_{a,b=0}^{p}
\Sigma_{i,j=p+1}^{9}\rightarrow T_{D(p-1)} \int d^{p}\sigma Tr
\Sigma_{a,b=0}^{p-1}
\Sigma_{i,j=p}^{9} + T_{D(1)} \int d^{2}\sigma Tr
\Sigma_{a,b=0}^{1}
\Sigma_{i,j=2}^{9} \nonumber \\
&& i\lambda F_{ap} \rightarrow \partial_{a} X^{p} \qquad i\partial_{p} X^{i} \rightarrow \frac{1}{\lambda}[X^{p},X^{i}]
\label{a6}
\end{eqnarray}

Replacing these equations in action (\ref{a5}) and doing some
algebra, we obtain :

\begin{eqnarray}
&& S_{Dp} =
-T_{Dp} \int d^{p+1}\sigma Tr
(\Sigma_{a,b=0}^{p}
\Sigma_{i,j=p+1}^{9}
\{\partial_{a}X^{i}\partial_{b}X^{i}-\frac{1}{2 \lambda^{2}}[X^{i},X^{j}]^{2}+\frac{\lambda^{2}}{4}
(F_{ab})^{2}
\})= \nonumber \\&& -T_{D(p-1)} \int d^{p}\sigma Tr
(\Sigma_{a,b=0}^{p-1}
\Sigma_{i,j=p}^{9}
\{\partial_{a}X^{i}\partial_{b}X^{i}-\frac{1}{2\lambda^{2}}[X^{i},X^{j}]^{2}+\frac{\lambda^{2}}{4}
(F_{ab})^{2}\})- \nonumber \\&& T_{D1} \int d^{2}\sigma Tr
(\Sigma_{a,b=0}^{1}
\Sigma_{i,j=2}^{9}
\{\partial_{a}X^{i}\partial_{b}X^{i}-\frac{1}{2\lambda^{2}}[X^{i},X^{j}]^{2}+\frac{\lambda^{2}}{4}
(F_{ab})^{2}\})- V_{separation}\rightarrow \nonumber \\&&
S_{Dp} = S_{D(p-1)} + S_{D1} - V_{separation}\rightarrow \nonumber \\&&
S_{Dp} + V_{separation} = S_{D(p-1)} + S_{D1}
\label{a7}
\end{eqnarray}

This equation shows that we  need to some extra energies that are applied for separating D1-brane from Dp-brane  $V_{separation} = - T_{D1} \int d^{2}\sigma Tr
\Sigma_{i,j=2}^{9}
\frac{1}{2\lambda^{2}}[X^{i},X^{j}]^{2}$. To supply this energy, we use of extra energies that are produced due to compactification of M0-branes.
 For this, we compactify $V_{Extra}$ on the circle with defining $<X^{p}> = \sqrt{\frac{T_{D1}}{12 T_{M0}\lambda^{2}}}$:

\begin{eqnarray}
&&  V_{Extra,1}= -6T_{M0}\int dt \Sigma_{M,N,L,E,F,G=0}^{9}\varepsilon_{MNLD}\varepsilon_{EFG}^{D}X^{M}X^{N}X^{L}X^{E}X^{F}X^{G}=\nonumber \\&&
- 6T_{M0}\int dt Tr(\Sigma_{M,N,E,F=0}^{9}\varepsilon_{MNpD}\varepsilon_{EFp}^{D}X^{M}X^{N}X^{p}X^{E}X^{F}X^{p} - \nonumber \\
&& 6T_{M0} \int dt \Sigma_{M,N,L,E,F,G=0,\neq p}^{8}\varepsilon_{MNLD}\varepsilon_{EFG}^{D}X^{M}X^{N}X^{L}X^{E}X^{F}X^{G} = \nonumber \\
&& - 6T_{M0}(\frac{T_{D1}}{12 T_{M0}\lambda^{2}})\int dt Tr(\Sigma_{M,N,E,F=0}^{9}\varepsilon_{MNpD}\varepsilon_{EFp}^{D}X^{M}X^{N}X^{E}X^{F} - \nonumber \\
&& 6T_{M0}\int dt \Sigma_{M,N,L,E,F,G=0,\neq p}^{8}\varepsilon_{MNLD}\varepsilon_{EFG}^{D}X^{M}X^{N}X^{L}X^{E}X^{F}X^{G} = \nonumber \\
&& - (\frac{T_{D1}}{2\lambda^{2}})\int dt Tr(\Sigma_{M,N=0}^{9}[X^{M},X^{N}]^{2}) - \nonumber \\
&& 6T_{M0} \int dt \Sigma_{M,N,L,E,F,G=0,\neq p}^{8}\varepsilon_{MNLD}\varepsilon_{EFG}^{D}X^{M}X^{N}X^{L}X^{E}X^{F}X^{G}= \nonumber \\
&& (\frac{T_{D1}}{2\lambda^{2}})\int dt Tr(\Sigma_{M,N=2}^{9}[X^{M},X^{N}]^{2}) - \nonumber \\
&& (\frac{T_{D1}}{2\lambda^{2}})\int dt Tr(\Sigma_{M,N=0}^{1}[X^{M},X^{N}]^{2})-  \nonumber \\
&&6T_{M0} \int dt \Sigma_{M,N,L,E,F,G=0,\neq p}^{8}\varepsilon_{MNLD}\varepsilon_{EFG}^{D}X^{M}X^{N}X^{L}X^{E}X^{F}X^{G}=\nonumber \\
&& V_{separation} + V_{Extra,2} + V_{Extra,3}
\label{a8}
\end{eqnarray}

where we define:
\begin{eqnarray}
&& V_{Extra,2}=-(\frac{T_{D1}}{2\lambda^{2}})\int dt Tr(\Sigma_{M,N=0}^{2}[X^{M},X^{N}]^{2}) \nonumber \\ && V_{Extra,3}=-6T_{M0} \int dt \Sigma_{M,N,L,E,F,G=0,\neq p}^{8}\varepsilon_{MNLD}\varepsilon_{EFG}^{D}X^{M}X^{N}X^{L}X^{E}X^{F}X^{G}
\label{a9}
\end{eqnarray}

These extra energies supply required energy for compactifying  D(p-1) on another circle. Following equation (\ref{a5}), we can write:

\begin{eqnarray}
&& S_{D(p-1)} = S_{D(p-2)} + S_{D1} - V_{separation}\rightarrow \nonumber \\&&
S_{D(p-1)} + V_{separation} = S_{D(p-2)} + S_{D1} \rightarrow \nonumber \\&& S_{Dp} + 2V_{separation} = S_{D(p-1)} + S_{D1} + V_{separation} = S_{D(p-2)} + S_{D1} + S_{D1}
\label{a10}
\end{eqnarray}

For example, by compactifying one D5-brane on two circles, one D3-brane and two D1-branes are created:

\begin{eqnarray}
&& S_{D(p-1)} = S_{D(p-2)} + S_{D1} - V_{separation}\rightarrow \nonumber \\&&
S_{D(p-1)} + V_{separation} = S_{D(p-2)} + S_{D1} \rightarrow \nonumber \\&& S_{D5} + 2V_{separation} = S_{D(4)} + S_{D1} + V_{separation} = S_{D(3)} + S_{D1} + S_{D1}
\label{a11}
\end{eqnarray}

This equation shows that the origin of D3-brane is D5-brane. On the hand, this D5-brane is produced by joining and growing D0-branes. Also, D0-branes are created by compactifying M0-branes. If we assume that our universe is located on D3-brane, we can claim that the main cause of the birth of universe is the transition of N M0-branes to a D3-brane via process $5 M0\rightarrow 5 D0 + \text{extra energy} \rightarrow D5 + \text{extra energy} \rightarrow D3 + D1 + D1$.

\section{ The Padmanabhan idea in D3-D1 system}\label{o2}
In this section, we will construct the Padmanabhan idea in D3-D1 system  and argue that the expansion of universe is controlled by the evolution of branes in extra dimensions. We will show that first D1-brane which is more close to our universe dissolves in it , increase inequality between number of degrees of freedom on the holographic surface and inside a bulk and leads to the inflation. Then, this brane gives it's energy to our universe brane, annihilates and inflation ends. After that, extra energies that are produced in  compactifications cause an expansion and deceleration epoch. Finally, another D1-brane interacts with our universe and leads to second phase of acceleration.

Using equation  (\ref{a5}), we can obtain relevant action for D3 and D1-branes:

\begin{eqnarray}
&& S_{D3}=-T_{D3} \int d^{4}\sigma Tr
(\Sigma_{a,b=0}^{3}
\Sigma_{i,j=4}^{9}
\{\partial_{a}X^{i}\partial_{b}X^{i}-\frac{1}{2 \lambda^{2}}[X^{i},X^{j}]^{2}+\frac{\lambda^{2}}{4}
(F_{ab})^{2}
\})
\label{a12}
\end{eqnarray}

\begin{eqnarray}
&& S_{D1} = -T_{D1} \int d^{2}\sigma Tr
(\Sigma_{a,b=0}^{1}
\Sigma_{i,j=2}^{9}
\{\partial_{a}X^{i}\partial_{b}X^{i}-\frac{1}{2 \lambda^{2}}[X^{i},X^{j}]^{2}+\frac{\lambda^{2}}{4}
(F_{ab})^{2}
\})
\label{a13}
\end{eqnarray}

Let us now to build the Padmanabhan idea in D3-D1 system. For this, we need to compute the contribution of  this system to the degrees of the surface and the bulk. To this end, we write the following relations between these degrees of freedom and the energy of D1 and D3-branes,

\begin{eqnarray}
&& N_{sur} \sim E_{D3} \qquad N_{bulk} \sim E_{D3-D1}= E_{D3}+E_{D1} \nonumber \\
&& N_{sur} - N_{bulk} \simeq E_{D1}
\label{a14}
\end{eqnarray}

where $E_{D3}$ and $E_{D1}$ are energies of D3 and D1-branes respectively.  Now, we want to calculate these energies by using the action (\ref{a12}):

\begin{eqnarray}
&& H_{D3}=\Sigma_{a,b=0}^{3}
\Sigma_{i,j=4}^{9}\Pi_{i} (\partial_{t}X^{i})-L_{D3}\qquad \Pi=\frac{\partial L}{\partial (\partial_{t}X^{i})}=-(\partial_{t}X^{i})\nonumber \\
&& L_{D3}=Tr
\Sigma_{a,b=0}^{3}
\Sigma_{i,j=4}^{9}
\{\partial_{a}X^{i}\partial_{b}X^{i}-\frac{1}{2 \lambda^{2}}[X^{i},X^{j}]^{2}+\frac{\lambda^{2}}{4}
(F_{ab})^{2}
\})\nonumber \\
&&E_{D3}=-T_{D3} \int d^{4}\sigma H=-T_{D3} \int d^{4}\sigma Tr
(\Sigma_{a,b=0}^{3}
\Sigma_{i,j=4}^{9}
\{\partial_{a}X^{i}\partial_{b}X^{i}+\nonumber \\
&&\frac{1}{2 \lambda^{2}}[X^{i},X^{j}]^{2}-\frac{\lambda^{2}}{4}
(F_{ab})^{2}
\})
\label{aa14}
\end{eqnarray}

\begin{eqnarray}
&& H_{D1}=\Sigma_{a,b=0}^{1}
\Sigma_{i,j=2}^{9}\Pi_{i} (\partial_{t}X^{i})-L_{D1}\qquad \Pi=\frac{\partial L}{\partial (\partial_{t}X^{i})}=-(\partial_{t}X^{i})\nonumber \\
&& L_{D1}=Tr
\Sigma_{a,b=0}^{1}
\Sigma_{i,j=2}^{9}
\{\partial_{a}X^{i}\partial_{b}X^{i}-\frac{1}{2 \lambda^{2}}[X^{i},X^{j}]^{2}+\frac{\lambda^{2}}{4}
(F_{ab})^{2}
\})\nonumber \\
&&E_{D1}=-T_{D1} \int d^{2}\sigma H=-T_{D1} \int d^{2}\sigma Tr
(\Sigma_{a,b=0}^{1}
\Sigma_{i,j=2}^{9}
\{\partial_{a}X^{i}\partial_{b}X^{i}+\nonumber \\
&&\frac{1}{2 \lambda^{2}}[X^{i},X^{j}]^{2}-\frac{\lambda^{2}}{4}
(F_{ab})^{2}
\})
\label{aaa14}
\end{eqnarray}

 Minimizing the relevant actions (\ref{a12}) and (\ref{a13}) and also energies (\ref{aa14}) and (\ref{aaa14}) for D3 and D1-branes yields the following condition \cite{q16}:

\begin{eqnarray}
&& \partial_{\sigma}X^{i}=\pm \frac{i}{2}\varepsilon^{ijk}[X^{j},X^{k}]
\label{a15}
\end{eqnarray}

The desired solution is given by:

\begin{eqnarray}
&& X^{i} = \pm \frac{\alpha^{i}}{2\sigma}
\label{a16}
\end{eqnarray}

where the $\alpha^{i}$ are an $N \times N$ representation of the SU(2) algebra,

\begin{eqnarray}
&& [\alpha^{i},\alpha^{j}] = i\varepsilon^{ijk}\alpha^{k}
\label{a17}
\end{eqnarray}

Using equation (\ref{a15}), we can obtain the minimum energy of D3 and D1-branes:

\begin{eqnarray}
&& E_{D3,min}=7T_{D3} \int d^{4}\sigma Tr
(\Sigma_{a,b=0}^{3}
\{\frac{\lambda^{2}}{4}
(F_{ab})^{2}
\})\nonumber\\&& E_{D1,min}=7T_{D1} \int d^{2}\sigma Tr
(\Sigma_{a,b=0}^{1}
\{\frac{\lambda^{2}}{4}
(F_{ab})^{2}
\})
\label{aa17}
\end{eqnarray}

Following rules in (\ref{a6}), we can obtain the solutions for  gauge fields in D3 and D1-branes:

\begin{eqnarray}
&&  \lambda F_{01} \text{ in D1-brane } \rightarrow \partial_{t} X^{1} \text{ in D3-brane} \qquad \nonumber \\&& F_{ab} \text{ in D3-brane } \rightarrow [X^{a},X^{b}]\text{ in D1-brane }\Rightarrow \nonumber \\&&
X^{i} \sim \frac{1}{2\sigma_{1}},\quad A^{i} \sim \frac{1}{2\sigma_{3}} \qquad \text{ in D1-brane }
\nonumber \\&& X^{i} \sim \frac{1}{2\sigma_{3}},\quad
A^{i} \sim  \frac{1}{2\sigma_{1}} \qquad \text{ in D3-brane }
\label{a18}
\end{eqnarray}

where $\sigma_{1}$ and $\sigma_{3}$ are coordinates of D1 and D3-branes respectively.
 With going time, $\sigma_{1}$ is decreased and reduced to zero at the end of inflation but $\sigma_{3}$ is increased.
  For this reason, we assume $\sigma_{1}\sim \frac{1}{t}$ and $\sigma_{3}\sim
  t$.  Choosing  these approximations needs some further discussion and explanations.
 According to this model, the universe is located on the D3-brane,
  thus due to  time  evolution and universe expansion, the D3-brane expands and $\sigma_{3}$
  which is the coordinate of D3-brane, grows and has a direct relation with time. On the other hand, the main
  cause of inflation is dissolving of D1-brane into D3-brane which represent our
  universe. Therefore, by passing time and universe inflation,
  $\sigma_{1}$, which is the coordinate of D1-brane, decreases and thus
  it is related with the inverse of time. Furthermore, due to the evolution in time
  and the disappearing of D1-brane, gauge fields which are stick to
  it, have to  vanish. As can be
  seen from equation (\ref{a18}), the gauge field on D1-brane
  is  related to $\displaystyle{\frac{1}{\sigma_{3}}}$ and thus by evolving  time
  and disappearing D1-brane and the gauge field, $\sigma_{3}$
   increases.

   With this assumption and using equations (\ref{aa14}), (\ref{aaa14})  and (\ref{a18}), and also condition in (\ref{a15}),  we can calculate energy of D1 and D3-branes and number of degrees of freedom on the holographic surface and one in a bulk:

\begin{eqnarray}
&& N_{sur} - N_{bulk} \simeq E_{D1} \simeq  14\pi^{2}l_{s}^{4} T_{D1}[\frac{t_{inf}-t}{t_{inf}}]\nonumber \\&&  N_{sur}\simeq E_{D3} + (E_{D1,inf}-E_{D1}) \simeq  \nonumber \\&& 15\pi^{2}l_{s}^{4} T_{D3}[\frac{t^{5}}{60}+\frac{t_{inf}t^{4}}{12}-\frac{t_{inf}^{2}t^{3}}{9}-\frac{t_{inf}^{3}t^{2}}{24}+\frac{t_{inf}^{4}t}{2}+ 14\pi^{2}l_{s}^{4} T_{D1}[\frac{t}{t_{inf}}]]
\label{a19}
\end{eqnarray}

where $(E_{D1,inf}=14\pi^{2}l_{s}^{4} T_{D1})$ is the energy of D1-brane at the beginning of inflation, $(E_{D1,inf}-E_{D1})$  is the amount of energy which dissolves in D3-brane and $t = t_{inf}$ is the time of end of inflation. As can be seen from this equation, difference between number of degrees of freedom on the holographic surface and bulk decreases with time and shrinks to zero at the end of inflation ($t= t_{inf}\Rightarrow N_{sur} = N_{bulk}$). This means that our calculations are consistent with the Padmanabhan idea and thus our model works.

These equations help us to obtain the relation between some of cosmological parameters like deceleration parameter and evolutions of D1 and D3-branes. To this end, first,  we  calculate Hubble parameter via following equation:

\begin{eqnarray}
&& N_{sur} = \frac{4\pi r_{A}^{2}}{l_{p}^{2}}\qquad r_{A}=\frac{1}{H}\Rightarrow \nonumber \\ && N_{sur} = \frac{4\pi }{(l_{p}H)^{2}} \simeq  15\pi^{2}l_{s}^{4} T_{D3}[\frac{t^{5}}{60}+\frac{t_{inf}t^{4}}{12}-\frac{t_{inf}^{2}t^{3}}{9}-\frac{t_{inf}^{3}t^{2}}{24}+\frac{t_{inf}^{4}t}{2}+ 14\pi^{2}l_{s}^{4} T_{D1}[\frac{t}{t_{inf}}]\Rightarrow \nonumber \\&& H=\frac{2}{\sqrt{15\pi l_{p}^{2}l_{s}^{4} T_{D3}[\frac{t^{5}}{60}+\frac{t_{inf}t^{4}}{12}-\frac{t_{inf}^{2}t^{3}}{9}-\frac{t_{inf}^{3}t^{2}}{24}+\frac{t_{inf}^{4}t}{2}]+ 14\pi^{2}l_{s}^{4} T_{D1}[\frac{t}{t_{inf}}]}}
\label{a20}
\end{eqnarray}

where H is the Hubble parameter and $l_{p}$ is the planck length. Using this equation, we can obtain deceleration parameter during inflation era in terms of time:

\begin{eqnarray}
&& q = \frac{d}{dt}(\frac{1}{H})-1 \simeq \sqrt{15\pi l_{p}^{2}l_{s}^{4} T_{D3}t_{inf}^{4}}[\frac{t^{4}}{24  t_{inf}^{4}}+\frac{t^{3}}{6 t_{inf}^{3}}-\frac{t^{2}}{6  t_{inf}^{2}}-\frac{t}{24 t_{inf}}]\nonumber \\&&-14\pi^{2}l_{s}^{4} T_{D1}t[1-\frac{t}{t_{inf}}]
\label{a21}
\end{eqnarray}

 This equation indicates that while the the age of universe (t) is increased, the deceleration parameter  reduces to lower negative values, turns over a minimum, increases and tends to zero at  $t = t_{inf}$ (see figure 1
). This means that the D1-brane is disappeared at the end of inflation, however the
 rate of acceleration of universe is increased very fast and tend to large values in this epoch.

\begin{figure*}[thbp]
\begin{tabular}{rl}
\includegraphics[width=10.0cm]{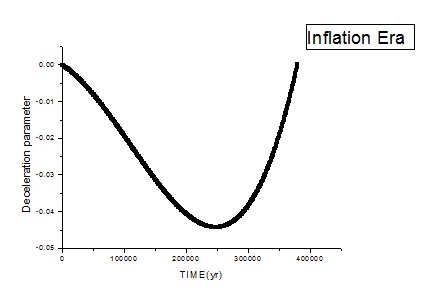}\\
\end{tabular}
\caption{  The deceleration  parameter  for inflation era
of expansion history as a function of the t where t is the age of
universe. In this plot, we choose $t_{inf}=380000$, $T_{D3}=10000.$ and $l_{s}=0.1$.}
\end{figure*}


With the disappearance of D1-brane, inflation ends and deceleration epoch begins. At this stage, extra energies that are produced in previous compactifications of branes cause to expansion of universe. Some of these energies are introduced in equation (\ref{a9}).  Applying $V_{Extra,2}=(\frac{T_{D1}}{2\lambda^{2}})\int dt Tr(\Sigma_{M,N=0}^{2}[X^{M},X^{N}]^{2})$ in this equation, assuming  $\sigma_{extra}\sim \frac{1}{t}$  and $\sigma_{3}\sim t$, and using equations (\ref{a12}), (\ref{a18}) and  (\ref{a15}),  we can obtain the amount of extra energy and also energy of D3-branes and number of degrees of freedom on the holographic surface and one in the bulk region:

\begin{eqnarray}
&& N_{sur} - N_{bulk} \simeq E_{extra,inf-t} \simeq -(\frac{T_{D1}}{2\lambda^{2}})\int dt Tr(\Sigma_{M,N=0}^{2}[X^{M},X^{N}]^{2}) \nonumber \\&& \simeq (\frac{T_{D1}}{4\pi^{2}l_{s}^{4}})[\frac{(t_{dec}-t_{inf})^{3}-(t-t_{inf})^{3}}{(t-t_{inf})^{3}}]\nonumber \\&& N_{sur}\simeq E_{D3} \simeq E_{D3,inf} + (E_{extra,inf-dec}-E_{extra,inf-t}) + 5T_{D3} \int d^{4}\sigma Tr
(\Sigma_{a=0}^{3}
\{\frac{\lambda^{2}}{4}
(\partial_{a}(\frac{1}{\sigma_{1}}))^{2}
\}) \simeq \nonumber \\&&  E_{D3,inf} + 15\pi^{2}l_{s}^{4} T_{D3}[\frac{(t_{dec}-t)^{5}}{60}+\frac{(t_{dec}-t_{inf})(t_{dec}-t)^{4}}{12}-\frac{(t_{dec}-t_{inf})^{2}(t_{dec}-t)^{3}}{9}-\nonumber \\&& \frac{(t_{dec}-t_{inf})^{3}(t_{dec}-t)^{2}}{24}+\frac{(t_{dec}-t_{inf})^{4}(t_{dec}-t)}{2}] + (\frac{T_{D1}}{4\pi^{2}l_{s}^{4}})[\frac{(t_{dec}-t_{inf})^{3}}{(t-t_{inf})^{3}}]
\label{a22}
\end{eqnarray}

where $E_{D3,inf}=\frac{480\pi^{2}l_{s}^{4} T_{D3}t_{inf}^{5}}{72}$ is the energy of D3-brane at the end of inflation, $(E_{extra,inf-dec}-E_{extra,inf-t})$ is the amount of energy that dissolves in D3-brane during deceleration era and $t = t_{dec}$ is the age of universe at the end of deceleration epoch. Similar to inflation era, difference between the number of degrees of freedom on the holographic surface and bulk decreases with time and shrinks to zero at the end of deceleration ($t= t_{dec}\Rightarrow N_{sur} = N_{bulk}$). The Hubble parameter during this era can be calculated as:

\begin{eqnarray}
&& N_{sur} = \frac{4\pi r_{A}^{2}}{l_{p}^{2}}\qquad r_{A}=\frac{1}{H}\Rightarrow \nonumber \\ && N_{sur} = \frac{4\pi }{(l_{p}H)^{2}} \simeq   E_{D3,inf} + 15\pi^{2}l_{s}^{4} T_{D3}[\frac{(t_{dec}-t)^{5}}{60}+\frac{(t_{dec}-t_{inf})(t_{dec}-t)^{4}}{12}- \nonumber \\&&\frac{(t_{dec}-t_{inf})^{2}(t_{dec}-t)^{3}}{9}- \frac{(t_{dec}-t_{inf})^{3}(t_{dec}-t)^{2}}{24}+\frac{(t_{dec}-t_{inf})^{4}(t_{dec}-t)}{2}]+ \nonumber \\&& (\frac{T_{D1}}{4\pi^{2}l_{s}^{4}})[\frac{(t_{dec}-t_{inf})^{3}}{(t-t_{inf})^{3}}]
\Rightarrow \nonumber \\&& H=\frac{4\pi}{l_{p}\sqrt{N_{sur}}}
\label{a23}
\end{eqnarray}

 With the help of this equation, we can derive deceleration parameter during deceleration epoch in terms of time:

\begin{eqnarray}
&& q = \frac{d}{dt}(\frac{1}{H})-1 \simeq \sqrt{15\pi l_{p}^{2}l_{s}^{4} T_{D3}(t_{dec}-t_{inf})^{4}}[-\frac{(t_{dec}-t)^{4}}{24  (t_{dec}-t_{inf})^{4}}-\frac{(t_{dec}-t)^{3}}{6 (t_{dec}-t_{inf})^{3}}\nonumber \\ &&+\frac{(t_{dec}-t)^{2}}{6  (t_{dec}-t_{inf})^{2}}+\frac{(t_{dec}-t)}{24 (t_{dec}-t_{inf})}]+ (\frac{T_{D1}(t-t_{inf})^{2}}{4\pi^{2}l_{s}^{4}})[1-[\frac{(t-t_{inf})^{2}}{(t_{dec}-t_{inf})^{2}}]]
\label{a24}
\end{eqnarray}

In figure.2, we present the deceleration parameter in terms of time. As can be seen from this figure, deceleration parameter  increases to higher  positive values, turns over a maximum, decreases and tends to zero at the end of deceleration epoch. Thus our model is consistent with previous predictions for deceleration era.

\begin{figure*}[thbp1]
\includegraphics[width=10.0cm]{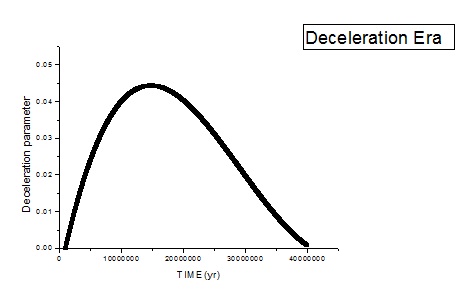}\\
\caption{  The deceleration  parameter  for deceleration era
of expansion history as a function of the t where t is the age of
universe. In this plot, we choose $ t_{inf}=380000$, $t_{dec}=40000000 yr$, $T_{D3}=10000.$ and $l_{s}=0.1$.}
\end{figure*}


After a period of time, second D1-brane become close to our universe D3-brane, dissolves in it and leads to present phase of acceleration. Similar to previous epochs, we assume that  $\sigma_{1}\sim \frac{1}{t}$  and $\sigma_{3}\sim t$, and use equations (\ref{a12}), (\ref{a13})  and (\ref{a18})and also condition in (\ref{a15}), to calculate the energy of D1 and D3-branes and number of degrees of freedom on the holographic surface and bulk:

\begin{eqnarray}
&& N_{sur} - N_{bulk}\simeq E_{D1,t-t_{dec}} \simeq 14\pi^{2}l_{s}^{4} T_{D1}[\frac{(t_{ac}-t_{dec})-(t-t_{dec})}{(t_{ac}-t_{dec})}]\nonumber \\&&   N_{sur}\simeq E_{D3} \simeq E_{D3,inf}+E_{D3,dec}+ (E_{D1,t_{ac}-t_{dec}}-E_{D1,t-t_{dec}}) +\nonumber \\&& 5T_{D3} \int d^{4}\sigma Tr
(\Sigma_{a=0}^{3}
\{\frac{\lambda^{2}}{4}
(\partial_{a}(\frac{1}{\sigma_{1}}))^{2}
\}) \simeq   E_{D3,inf} + E_{D3,dec} + 15\pi^{2}l_{s}^{4} T_{D3}[\frac{(t_{ac}-t)^{5}}{60}+\nonumber \\&&\frac{(t_{ac}-t_{dec})(t_{ac}-t)^{4}}{12}-\frac{(t_{ac}-t_{ac})^{2}(t_{ac}-t)^{3}}{9}- \frac{(t_{ac}-t_{dec})^{3}(t_{ac}-t)^{2}}{24}+\nonumber \\&&\frac{(t_{ac}-t_{dec})^{4}(t_{ac}-t)}{2}]-14\pi^{2}l_{s}^{4} T_{D1}[\frac{(t-t_{dec})}{(t_{ac}-t_{dec})}]
\label{a25}
\end{eqnarray}

In above equation, $E_{D3,dec}=\frac{480\pi^{2}l_{s}^{4} T_{D3}(t_{dec}-t_{inf})^{5}}{72}$ is the amount of energy that  D3-brane acquires during  deceleration era, $(E_{D1,t_{ac}-t_{dec}}-E_{D1,t-t_{dec}})$ is the amount of energy of D1-brane that dissolves in D3-brane during late time acceleration and $t = t_{ac}$ is the age of universe at the end of present acceleration epoch. Similar to previous epochs, difference between number of degrees of freedom on the holographic surface and bulk decreases with time and shrinks to zero at the end of late time acceleration ($t= t_{ac}\Rightarrow N_{sur} = N_{bulk}$). The Hubble parameter during this era can be obtained as:

\begin{eqnarray}
&& N_{sur} = \frac{4\pi r_{A}^{2}}{l_{p}^{2}}\qquad r_{A}=\frac{1}{H}\Rightarrow \nonumber \\ && N_{sur} = \frac{4\pi }{(l_{p}H)^{2}} \simeq  E_{D3,inf} + E_{D3,dec} + 15\pi^{2}l_{s}^{4} T_{D3}[\frac{(t_{ac}-t)^{5}}{60}+\frac{(t_{ac}-t_{dec})(t_{ac}-t)^{4}}{12}-\nonumber \\&&\frac{(t_{ac}-t_{ac})^{2}(t_{ac}-t)^{3}}{9}- \frac{(t_{ac}-t_{dec})^{3}(t_{ac}-t)^{2}}{24}+\frac{(t_{ac}-t_{dec})^{4}(t_{ac}-t)}{2}]-14\pi^{2}l_{s}^{4} T_{D1}[\frac{(t-t_{dec})}{(t_{ac}-t_{dec})}]\Rightarrow \nonumber \\&& H=\frac{4\pi}{l_{p}\sqrt{N_{sur}}}
\label{a26}
\end{eqnarray}

 Using this equation, we can calculate the deceleration parameter during present acceleration epoch in terms of time:

\begin{eqnarray}
&& q = \frac{d}{dt}(\frac{1}{H})-1 \simeq \sqrt{15\pi l_{p}^{2}l_{s}^{4} T_{D3}(t_{ac}-t_{dec})^{4}}[\frac{(t_{ac}-t)^{4}}{24  (t_{ac}-t_{dec})^{4}}+\frac{(t_{ac}-t)^{3}}{6 (t_{ac}-t_{dec})^{3}}\nonumber \\ &&-\frac{(t_{ac}-t)^{2}}{6  (t_{ac}-t_{dec})^{2}}-\frac{(t_{ac}-t)}{24 (t_{dec}-t_{inf})}]-14\pi^{2}l_{s}^{4} T_{D1}(t-t_{dec})[1-\frac{(t-t_{dec})}{(t_{ac}-t_{dec})}]
\label{a27}
\end{eqnarray}

 In figure.3, we show the deceleration parameter during late time acceleration era in terms of time. It is clear that deceleration parameter is negative at present stage which is a signature of acceleration. This result is in agreement with recent experimental data and thus our model works.

\begin{figure*}[thbp1]
\includegraphics[width=10.0cm]{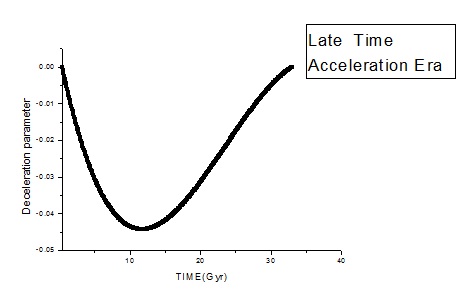}\\
\caption{  The deceleration  parameter  for acceleration era
of expansion history as a function of the t where t is the age of
universe. In this plot, we choose $ t_{ac}=33Gyr$, $t_{dec}=.4 Gyr$, $T_{D3}=10000.$ and $l_{s}=0.1$.}
\end{figure*}


\section{Summary and Discussion} \label{sum}
We constructed the Padmanabhan idea in M-theory  and argued that the birth and expansion of universe are controlled by the evolution of branes in extra dimensions. To this end,
first, we obtained the relevant action for N M0-branes by replacing Nambu-Poisson structure
of two dimensional manifolds in D-branes by the structure of three dimensional one. At second stage, we  compactified them on one circle and derived the action for N D0-brane. We showed that N D0-branes join to each other, grow and form a D5-brane. Then, we compactified this brane  on two circles and obtained the action for our universe-brane, two D1-branes and some extra energies that are created due to this compactifications. Next, we discussed that one of D1-branes which is more close to our universe-brane, dissolves in it, leads to an increase in difference between number of degrees of freedom on the holographic surface and bulk region and happening inflation era. After a short time, this D1-brane annihilates, the  number of degrees of freedom on the boundary surface and bulk region become equal, inflation ends and deceleration epoch begins. During this era, extra energies that are produced due to compactification are the main causes of expansion.  Finally, we argued that interaction of another D1-brane with our universe-brane leads to an inequality between degrees of freedom and occurring a new phase of acceleration.

\section*{Acknowledgments}
\noindent The work of Alireza Sepehri has been supported
financially by Research Institute for Astronomy and Astrophysics
of Maragha (RIAAM),Iran under research project No.1/4165-13.  We
thank of referee for nice comments that help us to improve our
paper.

 \end{document}